\documentclass[prl,twocolumn,showpacs,amsmath,amssymb]{revtex4}%
\usepackage{amsfonts}
\usepackage{bm}
\newcommand{\prodr}%
{\mathbin{\scriptstyle \bullet\mkern-3mu\bullet\mkern-3mu\bullet}}
\newcommand{\prodi}%
{\mathbin{\scriptstyle \circ\mkern-3mu\circ\mkern-3mu\circ}}
\newcommand{\prodra}%
{\mathbin{\scriptstyle \bullet\mkern-3mu\circ\mkern-3mu\circ}}
\newcommand{\prodrb}%
{\mathbin{\scriptstyle \circ\mkern-3mu\bullet\mkern-3mu\circ}}
\newcommand{\prodrc}%
{\mathbin{\scriptstyle \circ\mkern-3mu\circ\mkern-3mu\bullet}}
\newcommand{\prodia}%
{\mathbin{\scriptstyle \circ\mkern-3mu\bullet\mkern-3mu\bullet}}
\newcommand{\prodib}%
{\mathbin{\scriptstyle \bullet\mkern-3mu\circ\mkern-3mu\bullet}}
\newcommand{\prodic}%
{\mathbin{\scriptstyle \bullet\mkern-3mu\bullet\mkern-3mu\circ}}
\begin{document}
\title{Unification on gauge invariance and relativity}
\author{Haigang Lu}
\affiliation{College of Chemistry, Peking University, Beijing 100871, China, 
and Institute of Molecular Science, Shanxi University, Taiyuan, 
Shanxi 030006, China}
\email{luhg@pku.org.cn}
\altaffiliation[Present address: ]{Shanxi University.}
\date{\today}
%12.10.Dm Unified theories and models of strong and electroweak interactions
%04.20.Cv Fundamental problems and general formalism of relativity

\begin{abstract}
The inconsistence of Dirac-Weyl field equations with the universal $U(1)$
gauge invariance of neutrinos in quantum mechanics led to generalize the
special relativity to the generic relativity, which was composed of the
special relativity and its three analogues with different signatures of
metrics. The combination of the universal $U(1)$ gauge invariance and the
generic relativity naturally deduced the strong gauge symmetries $%
SU(3)$ and the electroweak ones $SU(2)\times U(1)$ in the Standard Model.
The universal $U(1)$ gauge invariance of elementary fermions is attributed
to its nature of complex line in three dimensional projective geometry.
\end{abstract}
\maketitle

The principle of gauge invariance plays a key role in the Standard Model
(SM), which is composed of Weinberg-Salam $SU(2)\times U(1)$ 
model\cite{Wein67,salam}
for the electroweak interactions and $SU(3)$ quantum 
chromodynamics\cite{gross,politzer}
for the strong interactions. Although the SM has achieved great success in
explaining experimental results in high energy physics\cite{pdg}, it has never
been accepted as a complete theory of fundamental physics. The ``minimal''
grand unified theories\cite{georgi}, in which the $SU(2)\times U(1)$ and $SU(3)$
gauge symmetries are incorporated into a simple $SU(5)$ one, was in
definite disagreement with the experimental limits on proton decay\cite{lang},
and the other ones must involve too many hypothetical particles to be
attractive\cite{grand}. On the other hand, the fundamental interactions 
in the SM
are mainly determined by the principle of gauge invariance and field
equations for free elementary fermions, i.e. Dirac-Weyl equations, which
come from quantum mechanics and the principle of relativity.

In quantum mechanics, the first postulate is that the state of a quantum
mechanical system is completely specified by a wavefunction $\psi(\bm{r},t)$ that
depends on the coordinates of the particle and on time,
which has the important property that $%
\psi^{\ast}(\mathbf{r},t)\psi(\mathbf{r},t)d\tau$ is the probability that the
particle lies in the volume element $d\tau$ located at $\mathbf{r}$ at time $%
t$. Any wavefunction can be multiplied by a phase factor $\exp(i\alpha)$,
where $\alpha$ is a scale function of coordinates and time, with no physical
consequence. The infinite set of these phase factors form the unitary group
called $U(1)$. An intrinsic and complete description of electromagnetism for
electron, one of elementary fermions, can be introduced by this $U(1)$ gauge
invariance. 
Because all other elementary fermions, neutrinos and quarks, are as fundamental
as electrons in quantum mechanics, it is reasonable
that every elementary fermions have a $U(1)$ gauge invariance,
called the \textit{universal} $U(1)$
gauge invariance for its universality for all elementary constituents of matter.
That is, it is necessary to
take the universal $U(1)$ gauge invariance as a fundamental principle in
the fundamental interactions in the SM.

In Weinberg-Salam model for electroweak interactions, the field equations for
all elementary fermions are the Dirac-Weyl equations\cite{Weyl29}, 
the only known field
equations for fermions (in units $\hbar=c=1$):%
\begin{equation}\label{weyleq}
\frac{\partial}{\partial t}\psi= \pm \left(  \sigma_{x}\frac{\partial}{\partial
x}+\sigma_{y}\frac{\partial}{\partial y}+\sigma_{z}\frac{\partial}{\partial
z}\right)\psi=\pm\bm{\sigma}\cdot\frac{\partial}{\partial\bm{r}}\psi,
\end{equation}
where $x$, $y$, $z$ and $t$ are spacetime coordinates and 
the $\sigma$'s are the $2\times 2$ Pauli spin matrices:
\[
\begin{array}{*{20}c}
   {\sigma _x  = \left[ {\begin{array}{*{20}c}
   0 & 1  \\
   1 & 0  \\
\end{array}} \right],} & {\sigma _y  = \left[ {\begin{array}{*{20}c}
   0 & { - i}  \\
   i & 0  \\
\end{array}} \right],} & {\sigma _x  = \left[ {\begin{array}{*{20}c}
   1 & 0  \\
   0 & { - 1}  \\
\end{array}} \right],}  \\
\end{array}
\]
$\bm{\sigma}=(\sigma_x,\sigma_y,\sigma_z)$ and $\bm{r}=(x,y,z)$ are
3-vectors.
The solutions of Eqs. (\ref{weyleq}) are two-component plane-wave solutions
\begin{equation}
\psi    = u_r \exp [ \pm i(\bm{p} \cdot \bm{r}- Et)],
\end{equation}
where $r=1,2$, and $u_r$'s are $2\times1$ constant matrices and will
be omitted for compactness later, $E$ is total energy and 
$\bm{p}=(p_x,p_y,p_z)$ is 3-momentum.
The Dirac-Weyl equations are invariant
under the $U(1)$ gauge transformation
\begin{subequations}
\begin{eqnarray}%
\psi=e^{\pm i(\bm{p}\cdot\bm{r}-Et)}&\rightarrow & \psi^{\prime}=
e^{\pm i(\bm{p}\cdot\bm{r}-Et\pm\alpha)}\\
 (\frac{\partial }{\partial t},\frac{\partial}{\partial \bm{r}}) &\rightarrow &
 (\frac{\partial}{\partial t} -i e\phi ,\frac{\partial}{\partial \bm{r}} 
 + ie\bm{A}) \\ 
 (\phi ,\bm{A}) &\rightarrow & (\phi  - \frac{\partial\alpha }{\partial t},
 \bm{A} + \frac{\partial \alpha }{\partial \bm{r}})  
\end{eqnarray}
\end{subequations}
where $e$ is the electronic charge of electron, 
$(\phi,\bm{A})= (\phi,A_{x},A_{y},A_{z})$ is a gauge potential,
$\alpha$ is a scale function of space and time.
In the Lagrangian of the $U(1)$ gauge theory, 
the only interaction term between elementary fermion field and the
gauge potential is 
\begin{equation}
ie\phi\psi^*\psi+ie\bm{A}\cdot \psi^*\bm{\sigma}\psi.
\end{equation}

In the first family of elementary fermions in the SM, 
the only observable ones are electron and neutrino.
The Dirac-Weyl equations correctly represented the electrons 
because its electronic charge is $e$ and 
the electromagnetic potential can be taken as the introduced gauge field. 
But it failed for the neutrinos whose electronic charge is zero
and there is no observable gauge field.
To find the correct field equations for neutrinos consistent with the 
universal $U(1)$ gauge
invariance, we must reconsider the kinematic part 
$(\frac{\partial}{\partial t},
\bm{\sigma}\cdot \frac{\partial}{\partial \bm{r}})$ 
of the Dirac-Weyl equations,
which comes from the energy-momentum, and therefore space-time,
relation of the special relativity.

From group theory,  
the Dirac-Weyl equations are the only field equations for
fermions in the special relativity.
We must go beyond the
special relativity within the general relativity to construct the correct 
field equations consistent with
the universal $U(1)$ gauge invariance.

The origin of the special relativity is 
its space-time relation%
\[
t^{2}=x^{2}+y^{2}+z^{2}.
\]
But if it is the only way for existence of ultimate constituents of matter or
elementary particles? In a remark concerning Weyl's 1918 work\cite{Weyl18},
 A. Einstein wrote that ``If
light-rays were the only means by which metrical relationships in the
neighbourhood of space-time point could be determined, ... .''\cite{Einstein}%
, which indicated that he took the light-rays as one of space-time relations,
not the only one. 

In four dimensional space-time, there are only four possible quadric
space-time relations (in the case of keeping the time direction):%
\begin{subequations}\label{relat}
\begin{eqnarray}
t^{2}=+x^{2}+y^{2}+z^{2}&=&\bm{r}\prodr \bm{r},\label{relata}\\
t^{2}=-x^{2}+y^{2}+z^{2}&=&\bm{r}\prodia \bm{r},\\
t^{2}=-x^{2}-y^{2}+z^{2}&=&\bm{r}\prodrc \bm{r},\\
t^{2}=-x^{2}-y^{2}-z^{2}&=&\bm{r}\prodi \bm{r},
\end{eqnarray}
\end{subequations}
in which each of operators $\bullet$ or $\circ$ between $\bm{r}$'s 
denotes that there is a extra factor $1$ or $i$ corresponding to one
component of space or momentum in the inner product of two 3-vectors. The only
differences between any two of these space-time relations are their 
signatures of
metrics. The relation Eq. (\ref{relata}) is just the space-time relation
invariant under Lorentz transformations of the special relativity, 
from which the
Dirac-Weyl equations for electrons were derived. The others could be taken as
the original space-time relations to derive the field equations for the other
elementary fermions: neutrinos and quarks. 

Because there are four species of space-time relations, there are four theories
of relativity keeping these quadric forms invariant. 
The whole set of these
theories of relativity is called \textit{generic relativity},
which is more general than the special relativity and still a special case of
the general relativity not including gravitations. 

Because of $-x^{2}=(ix)^{2}$,the field equations associated with quadric 
relations (\ref{relat})
are derived following the approach of Weyl\cite{Weyl29} or Lu\cite{lhg}, %
\begin{subequations}
\begin{eqnarray}
\frac{\partial}{\partial t}\psi &=&\pm\bm{\sigma}\prodr 
\frac{\partial}{\partial\bm{r}}\psi,
\label{fielda}\\
\frac{\partial}{\partial t}\psi &=&\pm\bm{\sigma}\prodia 
\frac{\partial}{\partial\bm{r}}\psi,
\label{fieldb}\\
\frac{\partial}{\partial t}\psi &=&\pm\bm{\sigma}\prodrc 
\frac{\partial}{\partial\bm{r}}\psi,
\label{fieldc}\\
\frac{\partial}{\partial t}\psi &=&\pm\bm{\sigma}\prodi
\frac{\partial}{\partial\bm{r}}\psi.
\label{fieldd}
\end{eqnarray}
\end{subequations}

The Eqs. (\ref{fielda}) is just the Dirac-Weyl equations for electron
whose electronic charge is $e$.
But what are the electronic charges for fermion fields represented by the other
field equations?

Firstly, in the Eqs. (\ref{fieldd}), which is symmetric in three dimensions of
space, its kinematic part (including the gauge field)
is $-(\frac{\partial}{\partial t}-ie\phi)\pm (i\bm{\sigma}\cdot\frac{\partial}
{\partial\bm{r}}+ie\bm{\sigma}\cdot\bm{A})$. 
Because not all four components of the gauge field are independent,
the temporal gauge $\phi=0$ can be chosen.
And in the same time, the space is chosen as real space in Mikowski spacetime
so that the only
interaction term is $\pm ie\bm{A}\cdot\psi^{\ast}i\bm{\sigma}\psi=
\mp e\bm{A}\cdot\psi^{\ast}\bm{\sigma}\psi$, 
which is anti-hermitian and unobservable
contrast with the electromagnetic interaction term 
$\pm ie\bm{A}\cdot\psi^{\ast}\bm{\sigma}\psi$ for electrons.
That is, no interaction
from the universal $U(1)$ gauge invariance is observed 
in the field equations (\ref{fieldd}) and
its (complex) charge is a pure imaginary charge $\pm ie$.
If the real part of this complex charge is interpreted as the electronic charge
of electromagnetic interactions, its electronic charge is zero. Then the
Eqs. (\ref{fieldd}) were proposed as the field equations for neutrinos.

Because of the isotrope of space in its three dimensions, we could 
formally take the complex
charge of the universal $U(1)$ gauge invariance as $1/3$ to each of 
space dimension.
From the analogical approach, it can be obtained that the complex charges are
formally $\pm(2/3+i/3)e$ and $\pm(1/3+2i/3)e$ and the absolute electronic 
charges are
$\pm 2e/3$ and $\pm e/3$ for the Eqs. (\ref{fieldb}) and (\ref{fieldc})
respectively. Then the Eqs. (\ref{fieldb}) and (\ref{fieldc})
were proposed as the field equations for quarks.

It is obvious that the Eqs. (\ref{fieldb}) and (\ref{fieldc}) are not symmetric
in the three space dimensions.
Because of the symmetry $SO(3)$ of space, 
each of them is
expanded into a set of equations respectively:%
\begin{subequations}\label{quarka}
\begin{eqnarray}
\frac{\partial}{\partial t}\psi &=&\pm\bm{\sigma}\prodia 
\frac{\partial}{\partial\bm{r}}\psi,\\
\frac{\partial}{\partial t}\psi &=&\pm\bm{\sigma}\prodib 
\frac{\partial}{\partial\bm{r}}\psi,\\
\frac{\partial}{\partial t}\psi &=&\pm\bm{\sigma}\prodic
\frac{\partial}{\partial\bm{r}}\psi,
\end{eqnarray}
\end{subequations}
and%
\begin{subequations}\label{quarkb}
\begin{eqnarray}
\frac{\partial}{\partial t}\psi&=&\pm\bm{\sigma}\prodra 
\frac{\partial}{\partial\bm{r}}\psi,\\
\frac{\partial}{\partial t}\psi&=&\pm\bm{\sigma}\prodrb 
\frac{\partial}{\partial\bm{r}}\psi,\\
\frac{\partial}{\partial t}\psi&=&\pm\bm{\sigma}\prodrc 
\frac{\partial}{\partial\bm{r}}\psi.
\end{eqnarray}
\end{subequations}

The solutions of the Eqs. (\ref{quarka}) and (\ref{quarkb}) are%
\[%
\left[\begin{array}
[c]{l}%
\exp[\pm i(\bm{p}\prodia\bm{r}-Et)]\\
\exp[\pm i(\bm{p}\prodib\bm{r}-Et)]\\
\exp[\pm i(\bm{p}\prodic\bm{r}-Et)]
\end{array}\right],\quad
%\]
%and
%\[%
\left[\begin{array}
[c]{l}%
\exp[\pm i(\bm{p}\prodra\bm{r}-Et)]\\
\exp[\pm i(\bm{p}\prodrb\bm{r}-Et)]\\
\exp[\pm i(\bm{p}\prodrc\bm{r}-Et)]
\end{array}\right],
\]
in which there
are two symmetries in each set of solutions: $U(1)^{3}$ from direct
product of the universal $U(1)$ gauge invariance 
and $SO(3)$ from the symmetry of space. The minimal group including
both of these symmetries as subgroups is $U(3)$, which can also be
deduced from the $U(3)$ symmetry of the 3-dimensional isotropic  
oscillator\cite{su} by analogism. 

From the decomposition $U(3)=SU(3)\times U(1)$, it is concluded that the
quarks must have a $SU(3)$ gauge symmetry, the origin of $SU(3)$
quantum chromodynamics for strong interactions, 
and an extra $U(1)$ one. 

Because there is no way to transform all the three space dimensions to real
ones synchronously in the Eqs. (\ref{quarka}) and (\ref{quarkb})
so that there is no free quark in our three
dimensional real space. It can be taken as the intrinsic origin of quark
confinement. 

Now, it can be concluded that the strong interactions can be described by a
secondary $SU(3)$ gauge theory on the universal $U(1)$ gauge invariance
and the generic relativity.

By far, there are eight solutions for all field equations from 
the generic relativity.
According to the real-imaginary symmetry of space, they can be put into four
two-dimensional representations of $U(2)$ gauge symmetry 
like the above $U(3)$ one.
\[
\begin{array}{*{20}c}
   {\left[ {\begin{array}{*{20}c}
\exp[\pm i(\bm{p}\prodr\bm{r}-Et)]\\
\exp[\pm i(\bm{p}\prodi\bm{r}-Et)]\\
\end{array}} \right],} & {\left[ {\begin{array}{*{20}c}
\exp[\pm i(\bm{p}\prodia\bm{r}-Et)]\\
\exp[\pm i(\bm{p}\prodra\bm{r}-Et)]\\
\end{array}} \right],}  \\
\\
   {\left[ {\begin{array}{*{20}c}
\exp[\pm i(\bm{p}\prodib\bm{r}-Et)]\\
\exp[\pm i(\bm{p}\prodrb\bm{r}-Et)]\\
\end{array}} \right],} & {\left[ {\begin{array}{*{20}c}
\exp[\pm i(\bm{p}\prodic\bm{r}-Et)]\\
\exp[\pm i(\bm{p}\prodrc\bm{r}-Et)]\\
\end{array}} \right].}  \\
\end{array}
\]
But this $U(2)$ symmetry have a difference with the usual U(2) group.
For instance, the basic vectors of the electron and neutrino are
$a=(-1,1,1,1)$ and $b=(-1,-1,-1,-1)$ in the space $(Et,p_xx,p_yy,p_zz)$.
The angle between these two vectors is $\arccos\frac{a\cdot b}{\sqrt{a\cdot a}
\sqrt{b\cdot b}}=\arccos(-1/2)= 2\pi/3$.
That is to say, this $U(2)$ gauge symmetry is deflective and the deflective
angle is $\theta= 2\pi/3-\pi/2=\pi/6$ with respect to the orthogonal case.
It is very interesting that
$\sin^{2}\theta=\sin^{2}(\pi/6)=0.25$, just like the Weinberg angle $\sin
^{2}\theta_{W}\thickapprox0.23$. The resulted gauge invariance is
$U(2)=SU(2)\times U(1)$, which just is the gauge symmetries
of the electro-weak interactions in the Weinberg-Salam model.

Now, it can be concluded that the electro-weak interactions can be described
by a secondary $SU(2)\times U(1)$ gauge theory on the universal $U(1)$
gauge invariance and generic relativity also.

Because the $SU(2)\times U(1)\times SU(3)$ Standard Model have been deduced from
the universal $U(1)$ gauge invariance and the generic relativity that is included
in the general relativity, there is only two universal interactions:
$U(1)$ gauge invariance and gravitations. 
It is well known that the gravitation could be attributed entirely to the
geometry of spacetime. What is the geometric significance of the universal 
$U(1)$ gauge invariance? 
In \cite{lhg}, the wavefunctions of the Dirac-Weyl equations are interpreted
as the homogeneous coordinates of lines in three dimensional projective geometry,
and it is easy to generalize to all cases in the generic relativity.
Then all wavefunctions of these elementary fermions can be interpreted as the 
homogeneous coordinates of lines 
in three dimensional projective geometry.
Each of these lines is a complex closed 1-dimensional geometric elements
and forms a $U(1)$ group.
That is, the linelike (non-pointlike) nature of elementary fermions 
is the geometric origin of the universal $U(1)$ gauge invariance
in microscopic world.

In conclusion, the universal $U(1)$ gauge invariance of elementary fermions
%from its line nature 
led to the special relativity to generalize to generic relativity,
and the Standard Model can be built as a secondary gauge theory on the 
universal $U(1)$ gauge invariance and the generic relativity.


\begin{thebibliography}{99}
\bibitem{Wein67} S. Weinberg, Phys. Rev. Lett. \textbf{19}, 1264(1967).
\bibitem{salam} A. Salam, in \textit{Elementary Particle Theory}, 
edited by N. Svartholm (Almqvist and Wiksell, Stockholm, 1969), 
p.367.
\bibitem{gross} D. J. Gross and F. Wilczek, Phys. Rev. D \textbf{8}, 3633(1973);
\textbf{9}, 980(1974).
\bibitem{politzer} H. D. Politzer, Phys. Rep. \textbf{14}, 129(1974). 
\bibitem{pdg} S. Eidelman et al., Phys. Lett. B \textbf{592}, 1 (2004).  
\bibitem{georgi} H. Georgi and S. L. Glashow, Phys. Rev. Lett. 
\textbf{32}, 438 (1974).
\bibitem{lang} P. Langacker, Phys. Rep., {\bf 72}, 185 (1980).
\bibitem{grand} R. N. Mohapatra, \textit{Unification and Supersymmetry},
3rd ed., (Springer-Verlag, New York, 2003).
\bibitem{Weyl29} H. Weyl, Z. Phys. \textbf{56}, 330(1929),
(an English translation is given in \cite{gauge}).
\bibitem{Weyl18} H. Weyl, Sitzungsber. Preuss. Akad. Berlin, 465(1918), 
(an English translation is given in \cite{gauge}).
\bibitem{Einstein} A. Einstein, Remark published at the end of \cite{Weyl18}, 
(an English translation is given in \cite{gauge}).
\bibitem{lhg} H. Lu, hep-ph/0312026.
\bibitem{su} J. M. Jauch and E. L. Hill, Phys. Rev. \textbf{57}, 641(1940).
\bibitem{gauge} L. O'Raifeartaigh, \textit{The dawning of gauge  theory},
(Princeton University Press, Princeton, 1997).
\end{thebibliography}
\end{document}